\documentclass{PoS}
\usepackage{wrapfig}
\usepackage{epsfig}
\usepackage{epstopdf}
\newcommand{\met}{\rlap{\,\,/}E_T}

\title{Status and Sensitivity Projections for the XENON100 Dark Matter Experiment}

\ShortTitle{Status and Sensitivity Projections for the XENON100 Dark Matter Experiment}

\author{Elena Aprile\thanks{Speaker (on behalf of the XENON100 Collaboration)}\\
        Columbia Astrophysics Laboratory, Columbia University, New York, NY 10027, USA\\
        E-mail: \email{age@astro.columbia.edu}}
     
\author{Laura Baudis\thanks{Speaker (on behalf of the XENON100 Collaboration)}\\
        Physik-Institut, Universit\"at Z\"urich, Z\"urich, 8057, Switzerland\\
        E-mail: \email{laura.baudis@physik.uzh.ch}}

\abstract{The XENON experimental program aims to detect cold dark matter particles via their 
elastic collisions with xenon nuclei in two-phase time projection chambers (TPCs). We are currently testing a new TPC at the 100 kg scale, XENON100. This new, ultra-low background detector, has a total of 170 kg of xenon (65 kg in the target region and 105 kg 
in the active shield). It has been installed at the Gran Sasso Underground Laboratory 
and is currently in commissioning phase. We review the design and performance of 
the detector and its associated systems, present status, preliminary calibration results, background prediction and projected sensitivity. With a 6000 kg-day background-free exposure, XENON100 will reach a sensitivity to spin-independent WIMP-nucleon cross section of $2\times10^{-45}$~cm$^2$ by the end of 2009. We also discuss our plan to upgrade the XENON100 experiment to improve the sensitivity by another order of magnitude by 2012.}

\FullConference{Identification of dark matter 2008\\
		 August 18-22, 2008\\
		 Stockholm, Sweden}

\begin{document}

\section{Introduction}

Numerous observations \cite{Spergel:2006hy} point to the existence of a non-luminous,
non-baryonic component of our universe known as dark matter.  The dark matter could be made of stable, massive,  
electrically and color neutral particles interacting very weakly with normal matter, generically referred to as Weakly Interacting Massive Particles  
(WIMPs). Direct detection of WIMPs is being pursued by several experiments placed in underground laboratories around the world. In recent years, experiments based on noble liquids have advanced at a faster rate than the more established cryogenic experiments.  In particular, the XENON10 experiment  has brought liquid xenon (LXe) detectors to the forefront of the field~\cite{Angle:2007uj,Angle:2008we}.

\begin{figure}[htbp]

\begin{center}
   \includegraphics[height=0.246\textheight]{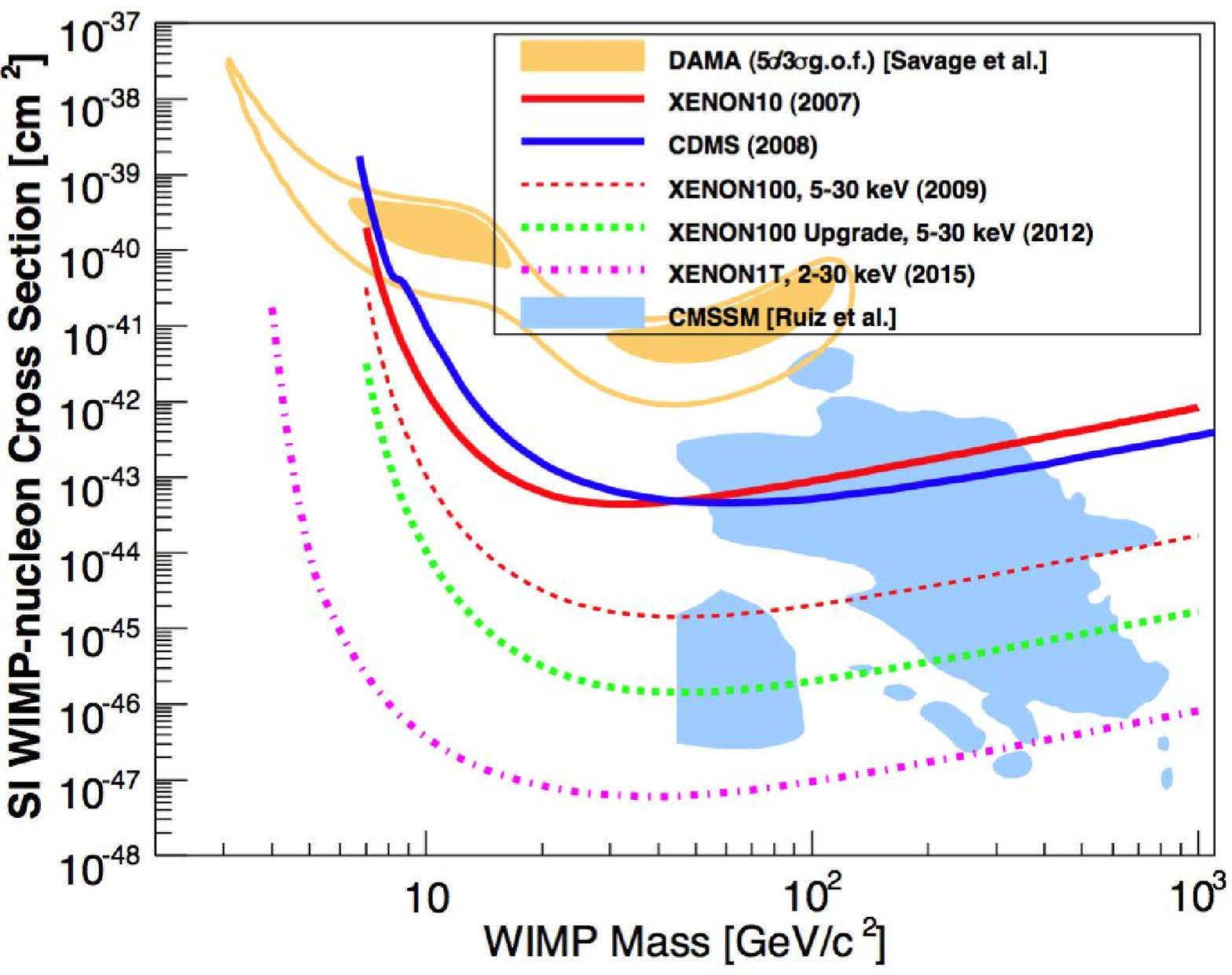}
   \includegraphics[height=0.24\textheight]{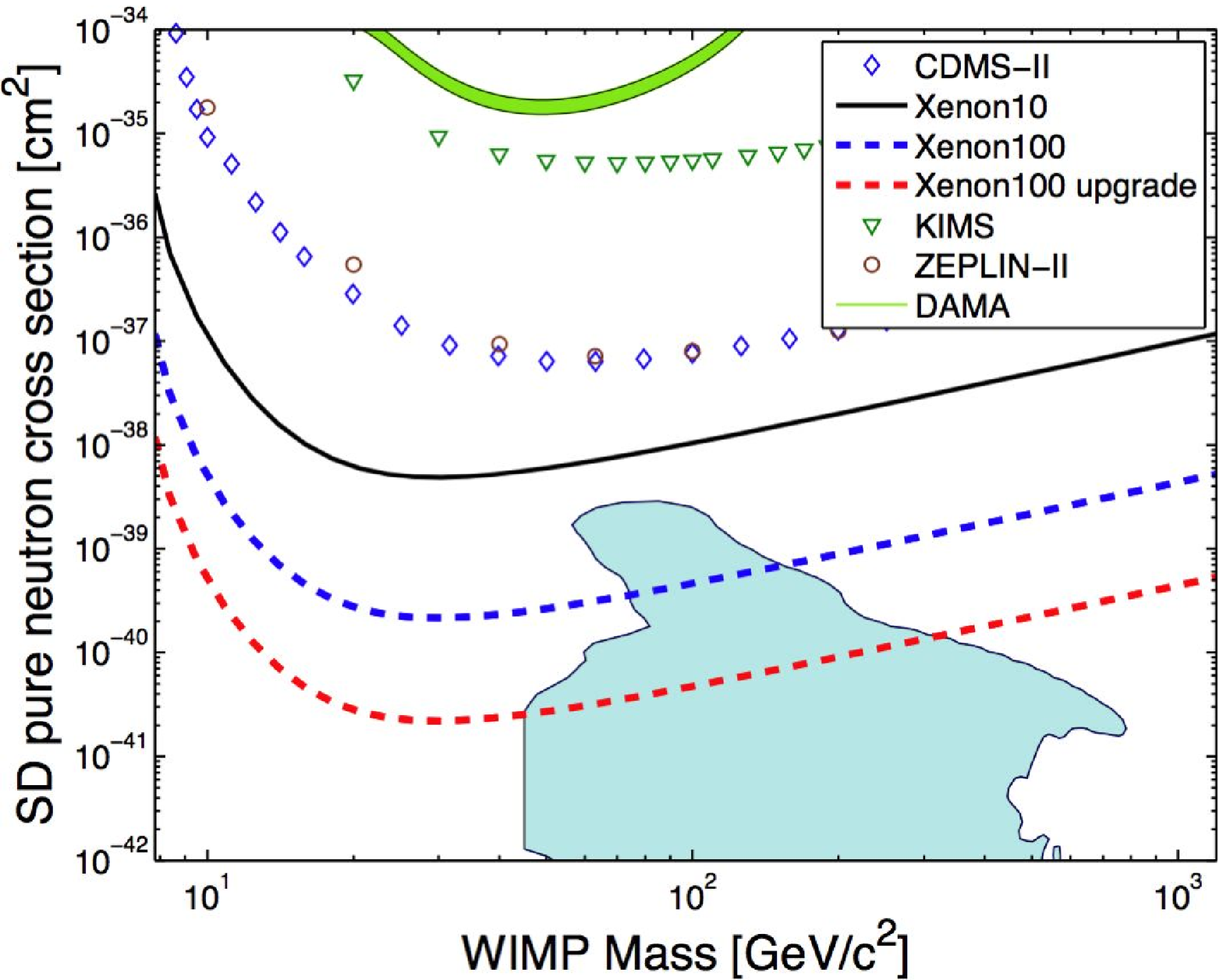}
\caption{\small Spin-independent (left) and spin-dependent pure neutron (right) WIMP-nucleon cross sections as a function of WIMP mass, probed by different stages of the XENON program. On the left, the low mass WIMP region allowed by DAMA~\cite{Bernabei:2008yi} is 
taken from \cite{Savage:2008er}, and the current most stringent limits from XENON10~\cite{Angle:2007uj} and CDMS~\cite{Ahmed:2008eu} are shown as solid curves. The blue shaded regions show the theoretical expectations within the CMSSM~\cite{Roszkowski:2007fd}. }\label{fig:exclusion}
\end{center}
\end{figure}

XENON10 was the first prototype developed within the XENON dark matter program to prove the concept of a two-phase xenon time projection chamber (XeTPC) for dark matter searches. The key technologies and performance characteristics relevant for the realization of a ton-scale experiment are being addressed with the current detector, XENON100, and its upgrade. The commissioning of XENON100 will be completed in early 2009; WIMP search data will be acquired throughout the year. With a raw exposure of  6000 kg-days, free of background events, XENON100 will reach a WIMP-nucleon cross section of $\sigma\sim2\times 10^{-45}$ cm$^2$ at a 
WIMP mass of 100 GeV/c$^2$.  An upgraded XENON100 will improve this sensitivity by another order of magnitude by 2012 (Figure~\ref{fig:exclusion}).

\section{The XENON100 Detector}

The XENON100 detector uses the same principle of operation and many design  features successfully tested in the XENON10 prototype~\cite{Xe10_Instr}. 
It is a position-sensitive XeTPC, with the sensitive LXe volume viewed by two arrays of photomultiplier tubes (PMTs), to detect simultaneously the primary scintillation signal (S1) and the ionization signal via the proportional scintillation mechanism (S2). The sensitive target can be ``fiducialized" to keep the inner core free of background. 
In addition, the high ionization density of nuclear recoils in LXe leads to an enhancement in S1 and reduction in S2, compared to electronic interactions. The S2/S1 ratio therefore discriminates WIMP nuclear recoils from $\gamma$ and $\beta$ backgrounds with an efficiency of 99.5--99.9\%, as demonstrated with XENON10~\cite{Angle:2007uj}.

\begin{wrapfigure}{r}{0.5 \textwidth}
\centering
\includegraphics[width=0.5\textwidth]{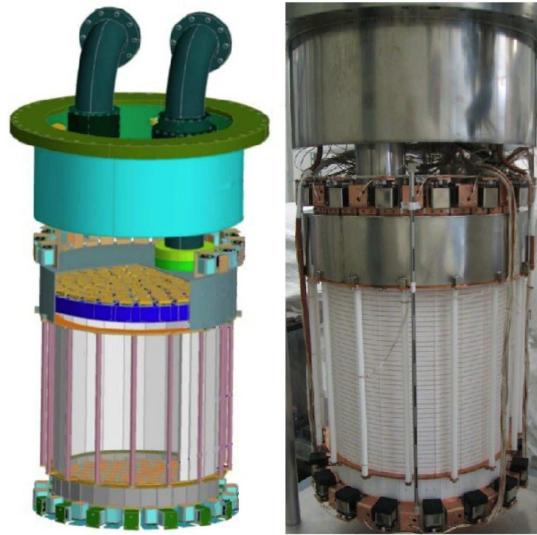}
\caption{\small Schematic view (left) and a picture (right) of the XENON100 time projection chamber.}\label{fig:xe100tpc}
\end{wrapfigure}

A schematic drawing of the XENON100 detector and a photo are shown in Figure~\ref{fig:xe100tpc}.
The active volume contains 65 kg of LXe and is instrumented with 178 PMTs in two arrays, 98 in the gas and 80 in the liquid. All PMTs are Hamamatsu R8520-06-Al 1\textquotedblright ~square, optimized for Xe 178 nm light and  selected for low radioactivity. The quantum efficiency (QE) of these tubes was recently improved from $\sim24$\% to $\sim35$\%.  

The active target  is enclosed in a PTFE cylinder of 15 cm radius and 30 cm height. It reflects scintillation light with high efficiency~\cite{Yamashita:2004}, and it optically separates the LXe target from the surrounding LXe, necessary to separate the TPC with its electric field from the walls of the vessel. 64 PMTs turn this outer LXe volume into an active LXe veto, with a total mass of 105 kg, including LXe layers above the top and below the bottom PMT arrays. 
Custom-made, low radioactivity, high voltage feedthroughs are used to bias the cathode and the anode, creating a 1~kV/cm drift field across the 30 cm LXe gap, and 13 kV/cm field in the 5~mm gas proportional scintillation region.  The LXe level is held stable within the stack of meshes, about 3 mm below the anode, using a construction similar to a diving bell, already used for XENON10. 
The top PMT array is mounted within the bell, a few centimeters above the meshes. 
The TPC structure, supported by the bell,  is enclosed in a double walled vessel made of SS 316Ti selected for its low activity, especially in  $^{60}$Co. 
All additional instrumentation was removed from the detector vessels as much as possible, and mounted outside the copper/lead/polyethylene shield, in particular all electrical feed-throughs and the cryocooler. This reduced the amount of radioactivity within the shield significantly.
The cooling system is based on a 170~W pulse tube refrigerator (PTR), originally developed for the MEG experiment~\cite{Haruyama:2005}. The PTR is used to liquefy Xe and to maintain the liquid temperature during operation.

\section{Xe Gas Handling, Purification and Kr Removal System}

The XENON100 detector requires a total of 170~kg of Xe to fill the target and the active veto. The gas is stored in four aluminum cylinders connected by high pressure valves, which can be cooled with LN$_2$ during recovery from the detector. 
To reduce electronegative impurities in commercial  Xe well below 1 part per billion (ppb) O$_2$ equivalent, required for long electron lifetime and long VUV photons absorption length, we are using the system developed for XENON10, based on continuous Xe gas circulation with purification through a high temperature metal getter (SAES). The $\sim$2~ms electron lifetime
demonstrated with XENON10 data~\cite{Xe10_Instr}, corresponds to four meters drift length, much longer than the 30 cm maximum drift in XENON100.

\begin{wrapfigure}{r}{0.4 \textwidth}
	  \centering
	  \includegraphics[height=0.4\textheight]{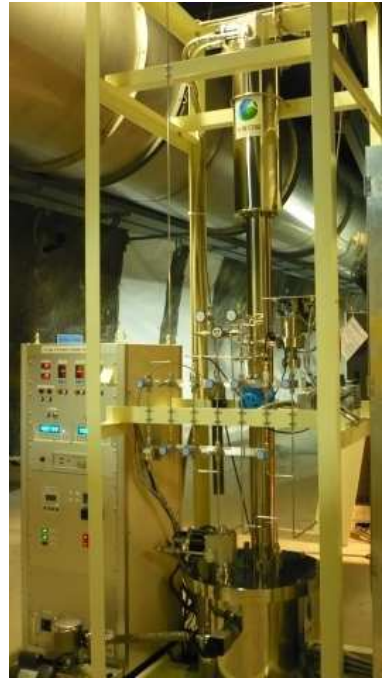}
\caption{\small Picture of the Kr distillation column during its commissioning at LNGS. }
   \label{fig:krcolumn}
\end{wrapfigure}

LXe, as a condensed noble gas, is readily purifiable for most radioactive impurities. The one notable exception is $^{85}$Kr,  present in commercial Xe gas at the ppm level. Beta decays of $^{85}$Kr (E$_{max}$=687\,keV, T$_{1/2}$=10.76 yr) present a serious background for a dark matter search. The gas used in XENON100 was  processed by the Spectra Gases Company to reduce the Kr concentration to $\sim$5~ppb, using their cryogenic distillation plant. This level has been achieved in XENON10~\cite{Xe10_Instr} with gas processed by the same company, and verified by positive identification of beta-gamma coincidences with 1.46~$\mu$s time difference from $^{85}$Kr$(\beta)\rightarrow^{85m}$Rb$(\gamma)\rightarrow^{85}$Rb decays  
 at a 0.454\% branching ratio.  During the first background run with  XENON100, we identified these "delayed-coincidence" events and inferred a Kr/Xe level of  7~ppb, consistent with the value quoted by Spectra Gases. 
 In order to reduce the $^{85}$Kr level to $<$50~ppt, required by the XENON100 sensitivity goal (50~ppt of $^{85}$Kr yield $10^{-3}$ evts/kg/keV/day), we have purchased a cryogenic distillation column made by Taiyo-Nippon Sanso (shown in Figure~ \ref{fig:krcolumn}). The column has been commissioned at LNGS and has been used to purify part of the XENON100 gas. It is 3~m tall and is designed to deliver a factor of 1000 reduction in Kr at a purification speed of 0.6~kg/hour. \vspace{0.5cm}

\section{Electronics and Data Acquisition }

The XENON100 data acquisition (DAQ) system generates the trigger for the TPC, 
digitizes the waveforms of the 242 PMTs, and stores the data in an efficient way.
 The PMT signals are first amplified by a factor 10 (Phillips 776 amplifiers), and then
digitized by CAEN V1724 Flash ADCs with 10 ns sampling period, 14 bit resolution, and 40 MHz
bandwidth

We have implemented an algorithm on the on-board FPGA of the ADC channels to digitizes the waveform only within an adjustable time window around a signal exceeding an adjustable threshold. This reduces the event size by 90--95\% and allows  full acquisition at a speed up to 50 Hz, sufficient for taking calibration data and much larger than the total background rate in the detector ($\sim$1~Hz). 
The complete XENON100 DAQ system is installed underground and is being used for measurements since several months. 

\section{Initial Results from XENON100}

At the time of this writing the detector is filled with 150 kg of Xe (the remaining 20 kg will be added at the end of November) which we continue to purify with the closed circulation system to improve light and charge yields.

\begin{figure}[htbp]
   \centering
   \includegraphics[scale=0.3]{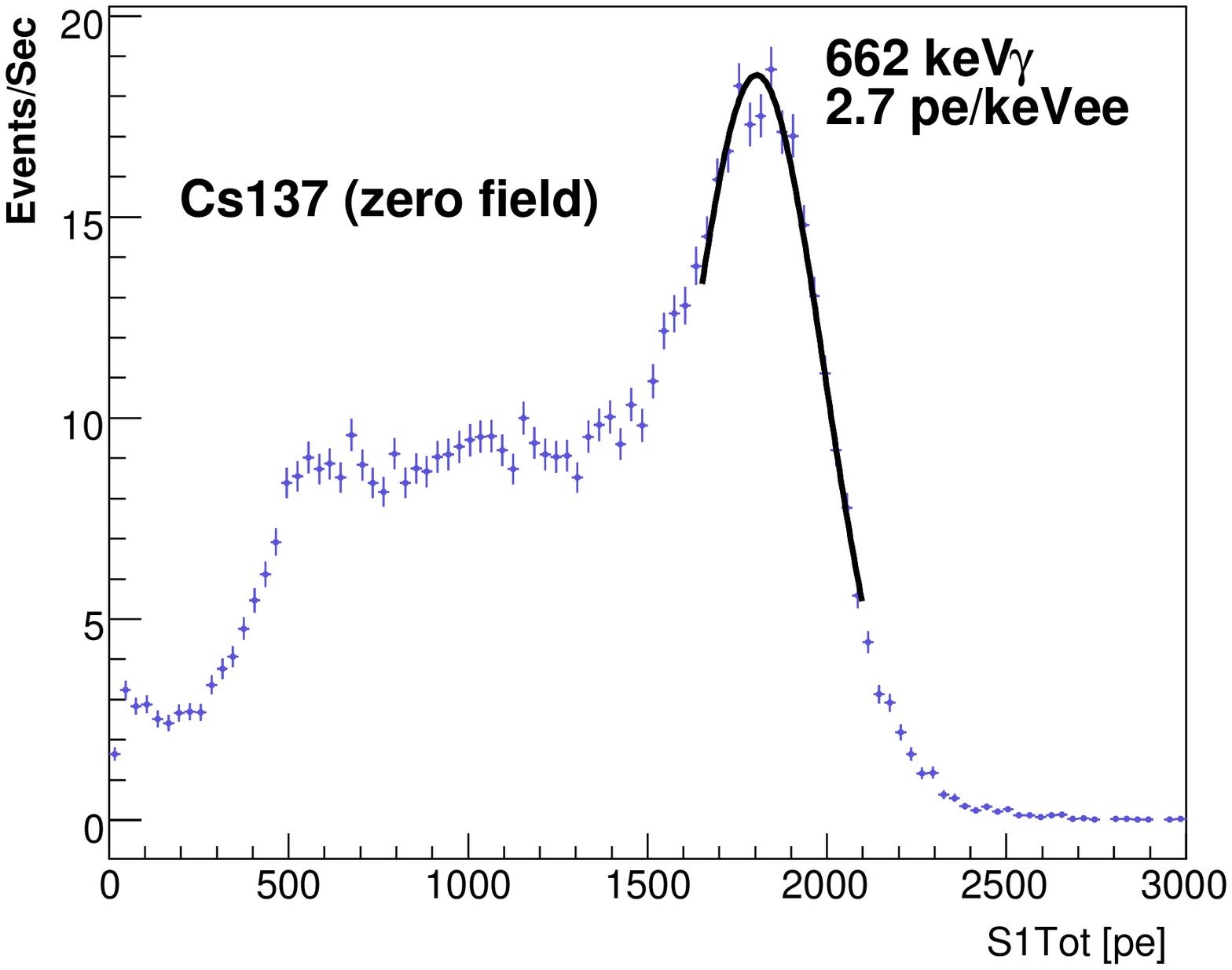} 
   \includegraphics[scale=0.5]{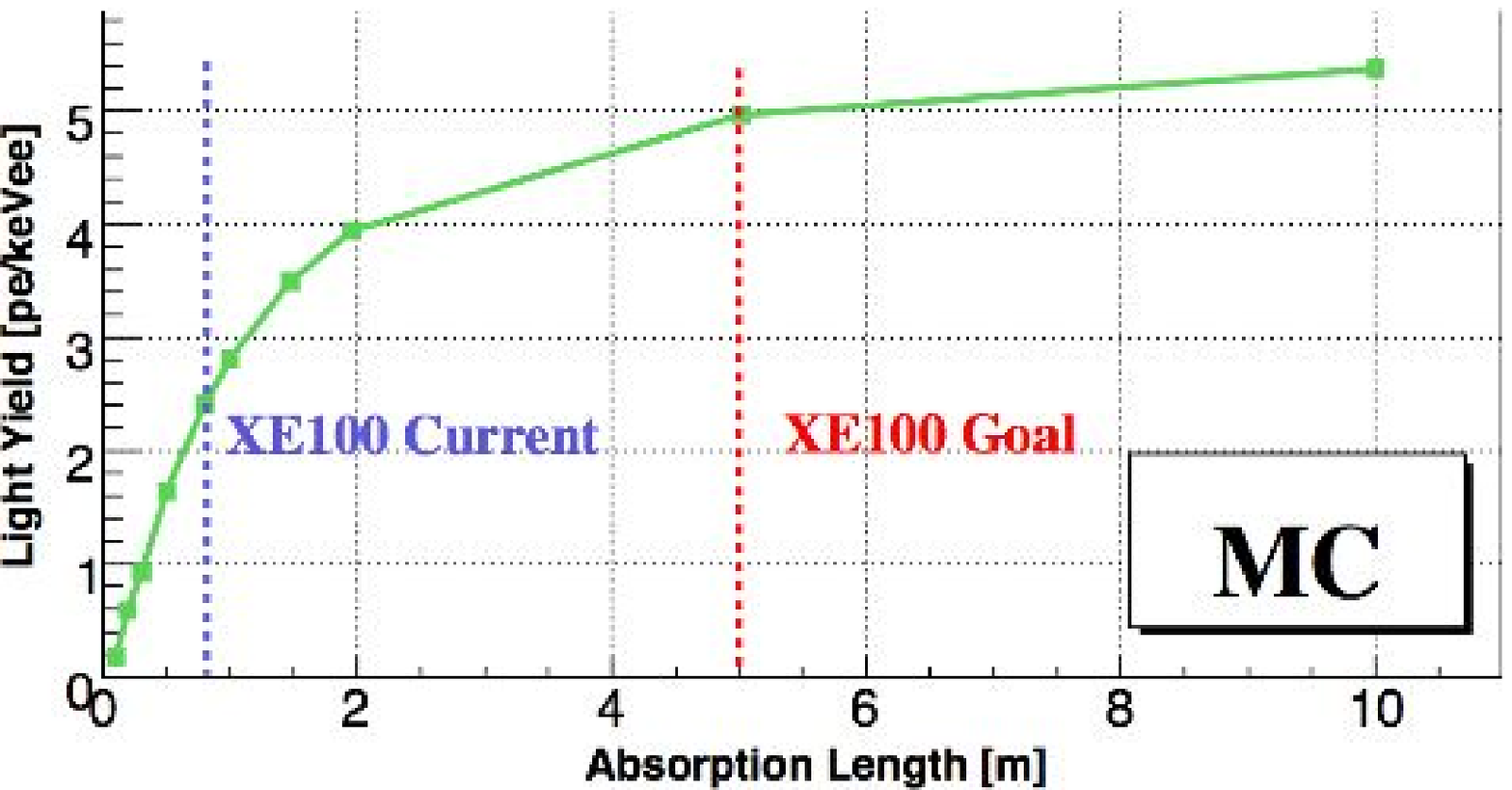} 
   \caption{\small (Left) The measured scintillation light spectrum based on S1 signals from $^{137}$Cs. (Right) Current and projected goal of the S1 light yield by increasing the LXe purity in XENON100.}
   \label{fig:lightyield}
\end{figure}

The S1 and S2 response of the XENON100 TPC is  monitored with external gamma sources ($^{57}$Co, $^{137}$Cs, $^{60}$Co and $^{228}$Th). We are investigating a new calibration source, $^{83m}$Kr, which is a decay product of $^{83}$Rb, has a short half-life of 1.83\,h and decays via a cascade of 32 and 9.4~keV transitions. 
$^{83m}$Kr will be regularly introduced in the TPC, allowing S1 and S2 calibrations uniformly throughout the volume. The neutron calibration will occur in the middle of the WIMP search run to verify the TPC's response to nuclear recoils (NRs). Simulation studies show that a one day calibration with a 220~n/s AmBe source will produce more than 1000 NR~events/kg in the central 50~kg target in the WIMP search energy window. 

\begin{wrapfigure}{r}{0.5 \textwidth}
 	\centering
   \includegraphics[scale=0.45]{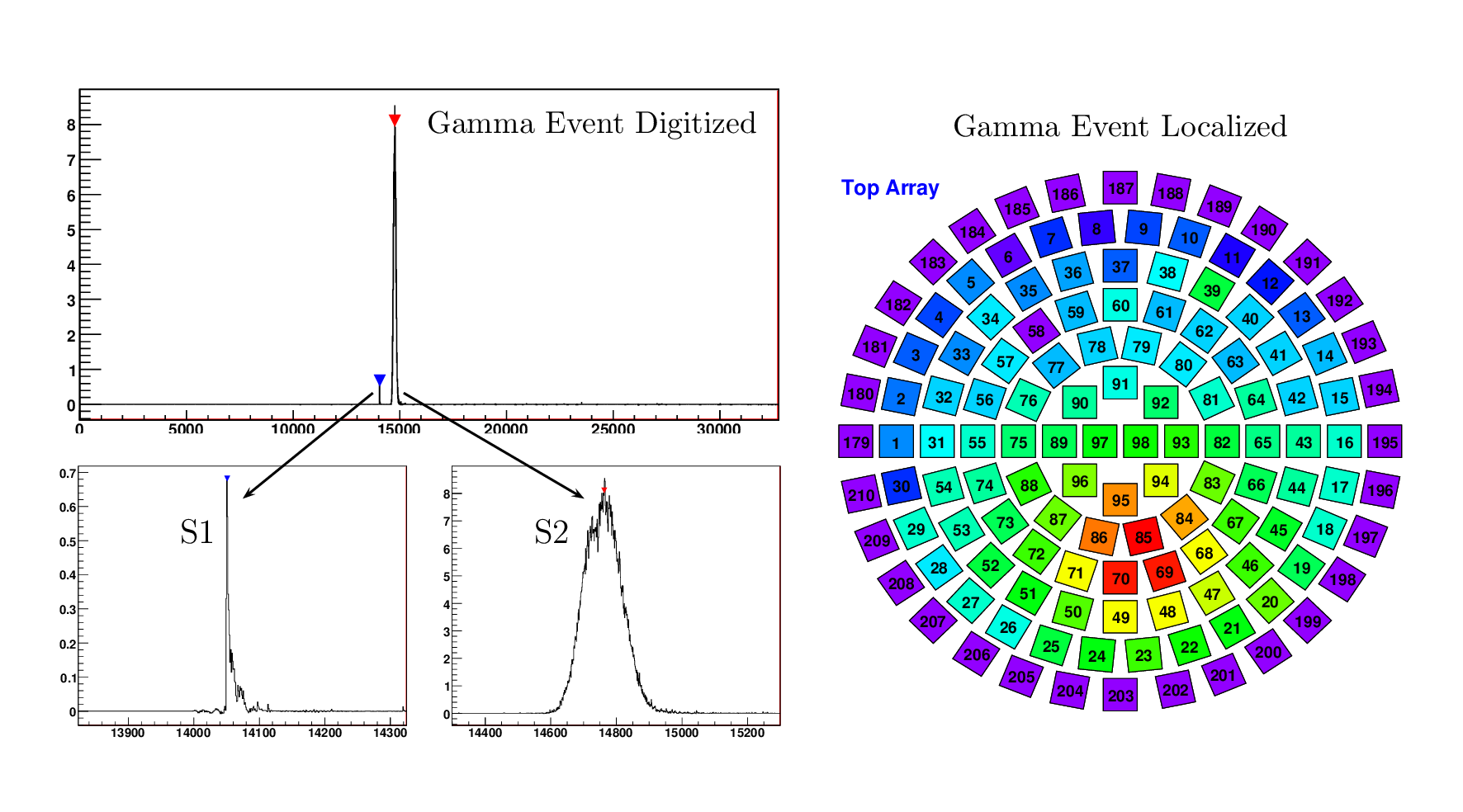} 
   \caption{\small (Left) A typical event waveform from XENON100, showing both S1 and S2 signals. (Right) S2 hit pattern on the top PMT array for  the same event.}
   \label{fig:s2s1_wave}
\end{wrapfigure}

Figure~\ref{fig:lightyield} shows the current S1 light spectrum of $^{137}$Cs 662\,keV gamma-rays, measured in XENON100, showing a light yield of 2.7\,pe/keVee (zero field), limited by the presence of impurities, mostly $\rm H_2O$. By comparing with the MC studies of the light collection efficiency, the current light yield implies a 1\,m absorption length of UV photons in LXe. With continuous circulation and purification of the Xe through the high temperature getter, we expect to achieve a S1 light yield of at least 5\,pe/keVee (zero field) at a 5\,m absorption length, a value obtained in XENON10~\cite{Xe10_Instr}. This yields an average S1 signal larger than 3\,pe for 5\,keV nuclear recoils, based on a recent measurement of the relative scintillation efficiency $L_{eff}$~\cite{Aprile:2008rc}, allowing $\sim$100\% detection efficiency for NRs as low as 5\,keV.

Two-phase operation of the TPC has started. Figure~\ref{fig:s2s1_wave} (left) displays a waveform from the summed signals, showing clear S1 and S2 signals from a single scatter electron recoil (ER) event. Calibration data from external gamma ray sources ($^{137}$Cs) are continuously collected to monitor the electron lifetime in the LXe. The top PMT array was designed to optimize the radial position resolution, allowing efficient fiducial volume cuts. Figure~\ref{fig:s2s1_wave} (right) shows a typical S2 hit pattern on the top PMT array from the same gamma event. From the S2 hit pattern, $XY$ positions can be reconstructed with a radial position resolution of about 2\,mm from a few keVee ER events, due to the fine granularity of the 1" PMTs.

\begin{figure}[htbp]
   \centering
    \includegraphics[scale=0.35]{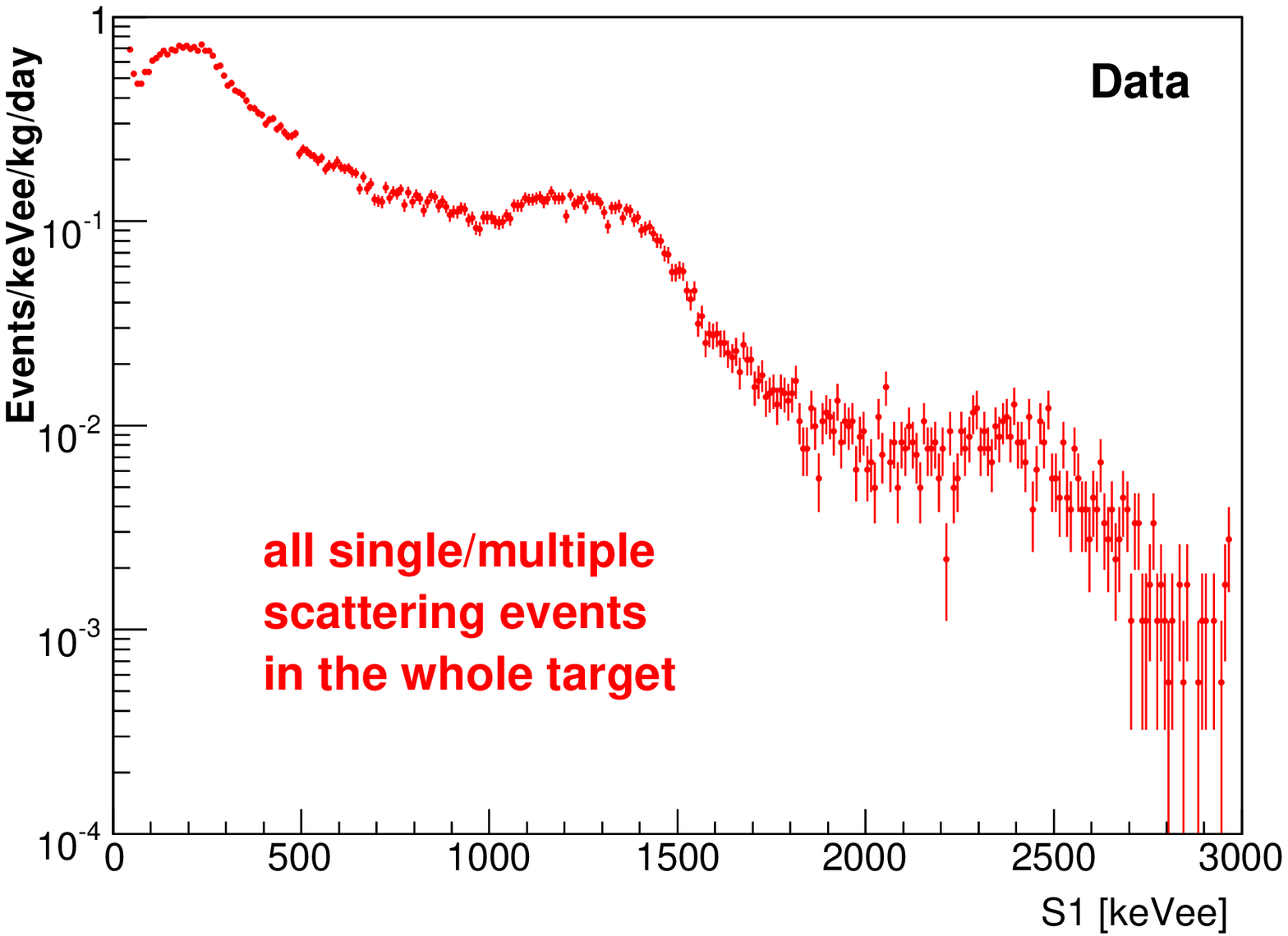}
    \includegraphics[scale=0.35]{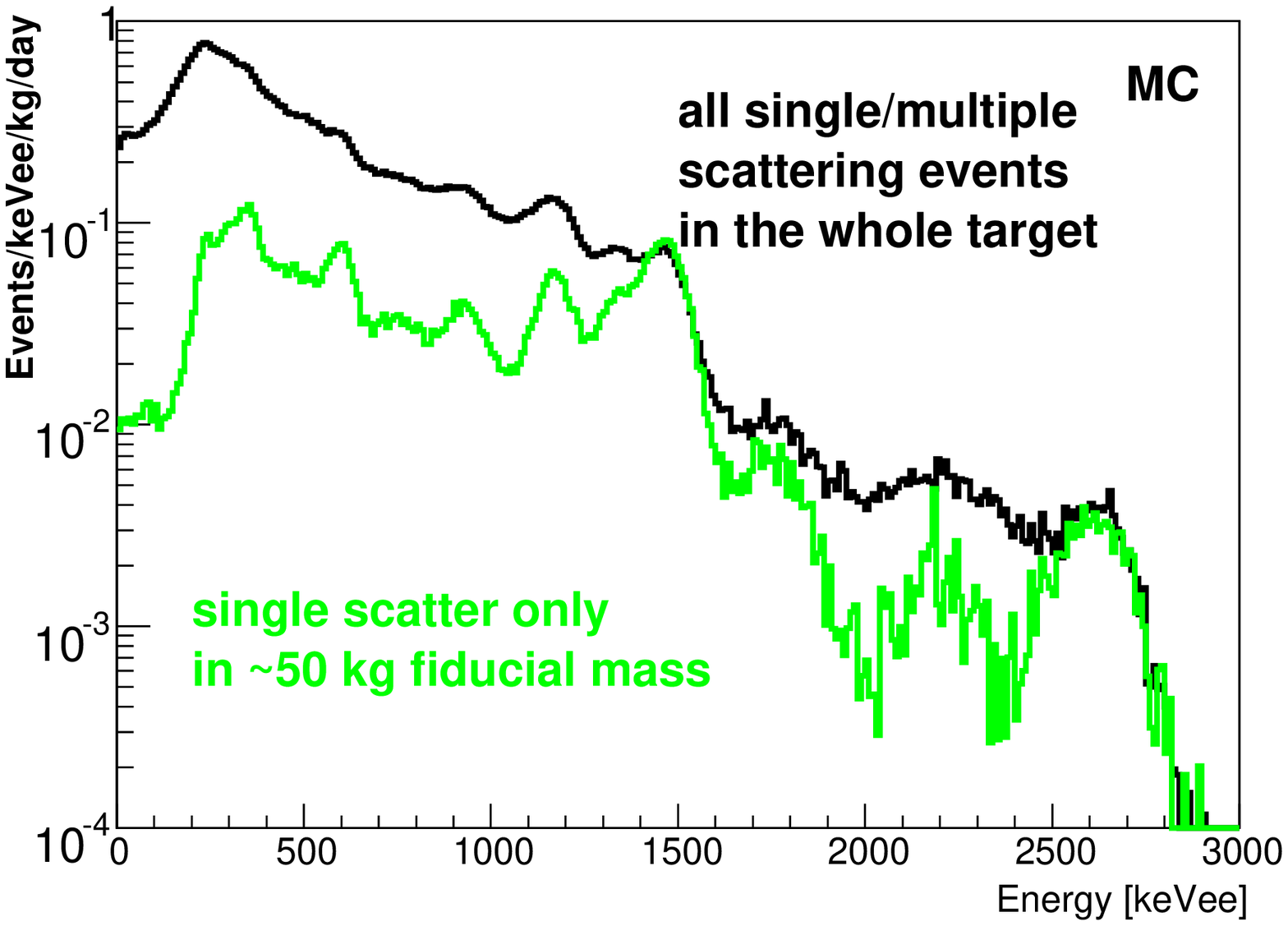}
   \caption{\small Measured (left) and MC predicted (right) background spectra in XENON100.}
   \label{fig:bkg_s1}
\end{figure}

The first background measurement in XENON100 has been performed with the S1 signal. The measured spectrum, shown in Figure~\ref{fig:bkg_s1}, is consistent with Monte Carlo predictions (see Section~\ref{sec:bkg}). 

\section{The XENON100 Upgrade}

LXe is an excellent medium
to shield the fiducial volume from penetrating gamma-rays from radioactive decays of 
U/Th, Co, and K. One may thus rely on the position sensitivity
of a TPC to cut the outer volume and achieve a background free inner
target. However, LXe is expensive and larger TPCs become increasingly complex:
increased drift length requires higher voltage to be brought into the detector and
shielded from the photosensors; purification becomes more demanding; event rates
in the TPC need to be reduced to accommodate longer drift times, etc. Moreover,
self-shielding is less effective for neutrons. It is therefore more cost-effective
to achieve a high ratio of \textit{fiducial/total} volume, for a given
sensitivity goal, by decreasing the radio-purity of the photosensors and the cryostat.

In the upgraded XENON100 detector the current bottom R8520 PMT array will be 
replaced with with 19 QUPID 
(Quartz Photon Intensifying Detectors) sensors \cite{QUPIDs}. We will simultaneously remove most of the activity from the
stainless steel (SS) cryostat by replacing it with a low background, 
oxygen free copper (OFHC) one. As the activity of the top PMTs will dominate the 
background, we will double the drift length from current \mbox{30 cm} to 
\mbox{60 cm}. The activity from the top cryostat assembly, made of the lowest activity SS
we have identified so far, will add negligible background compared to the top
PMTs. 

\begin{wrapfigure}{r}{0.6\textwidth}
\includegraphics[width=\linewidth,clip]{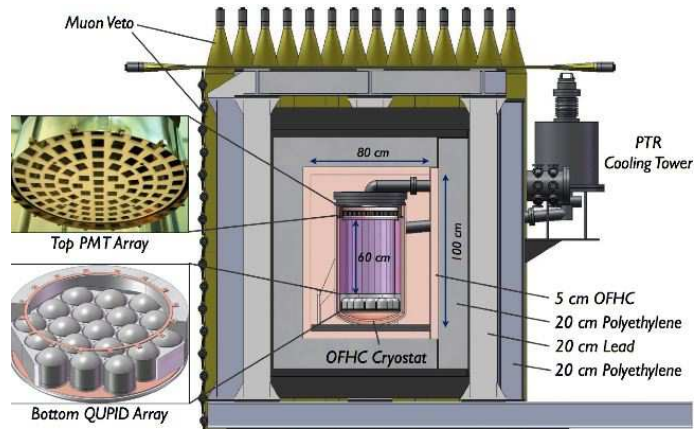}
\caption{\small The upgraded XENON100 with QUPID sensors on the bottom and an OFHC Cu cryostat in the improved shield.}
\label{fig:xe100_tpc}
\end{wrapfigure}

The upgraded XENON100 TPC will not only feature greatly reduced background but
will  be a stepping stone for a future ton-scale XeTPC: it will validate the
new QUPID technology, and  test the HV handling and long drift in LXe, both essential for a XENON1T.  At the same time, the current shield needs to be improved by adding 20~cm of polyethylene, mounted to the existing steel structure, for additional neutron moderation. Outside of the moderator, we will place a muon veto to tag events caused by high energy neutrons from muon spallation in the shield and other passive materials. The muon veto is made of 60
panels of polished Bicron BC408 plastic scintillator, optically coupled to two
Hamamatsu 5946 PMTs, one on each side. The muon veto will cover all
sides of the shield except the bottom, with an estimated efficiency at about 98\%. A cross-sectional view of the proposed detector mounted inside the upgraded shield is shown in Figure~\ref{fig:xe100_tpc}.  

\section{Background and Sensitivity Projections}
\label{sec:bkg}

In the following, we summarize the estimated background contributions from different sources in the current and the upgraded XENON100 detectors, based on detailed Monte Carlo simulations and the measured radioactivity of detector construction materials and components. The materials used in the XENON100 TPC and shield were carefully selected for low intrinsic radioactivity (Table~\ref{tab:screen_results}) with a dedicated screening facility consisting of a 2.2~kg high purity Ge detector in an ultra-low background Cu cryostat and Cu/Pb shield, operated at LNGS.

\begin{table}[h!]
\centering
{\footnotesize
\begin{tabular}{lccccccc}
\hline
\hline
 & Unit & Quantity & $^{238}$U & $^{232}$Th & $^{40}$K & $^{60}$Co  & $^{210}$Pb\\
\textit{TPC Material} & & used & [mBq/unit] &  [mBq/unit] &  [mBq/unit] &  [mBq/unit] & [Bq/unit]\\
\hline
R8520 PMTs & PMT & 242  &  0.15$\pm$0.02	& 0.17$\pm$0.04	& 9.15$\pm$1.18 & 1.00$\pm$0.08 & \\
PMT bases & base &  242 & 0.16$\pm$0.02	& 0.07$\pm$0.02	& $<$ 0.16	& $<$ 0.01 & \\
Stainless steel & kg & 70  & $<$ 1.7	&  $<$ 1.9	& $<$ 9.0	& 5.5$\pm$0.6 &\\
PTFE & kg & 10 & $<$ 0.31	&  $<$ 0.16	&  $<$ 2.2	&  $<$ 0.11 &\\
QUPID  & QUPID   &   -  &  $<$0.49  & $<$0.40   &  $<$2.4  & $<$0.21 & \\
\hline
\textit{Shield Material} & & & & & & & \\
\hline
Copper  & kg & 1600 & $<$ 0.07 &  $<$ 0.03	& $<$0.06 &  $<$0.0045 &\\ 
Polyethylene & kg & 1600  & $<$ 3.54 &	 $<$ 2.69 &  $<$ 5.9 	&  $<$ 0.9 & \\
Inner Pb (5 cm)  & kg & 6300 & $<$ 6.8	&  $<$ 3.9	&    $<$ 28 & $<$ 0.19 & 17$\pm$5	\\
Outer Pb (15 cm) & kg & 27200  & $<$ 5.7	&  $<$ 1.6	&   14$\pm$6 & $<$ 1.1 & $516\pm90$	\\
\hline
\hline
\end{tabular}}
\caption{\small Radioactivity of XENON100 materials: average values are given if different activities were obtained for different material samples, such as 
different batches of PMTs and stainless steel. Upper limits are given if no activity above background was found. Radioactivity from other components, 
such as screws and cables, are negligible (at least a factor of 10 lower compared to those in the table).}
\label{tab:screen_results}
\end{table}

\subsection{Background prediction}

\begin{figure}
\centering
\includegraphics[scale=0.3]{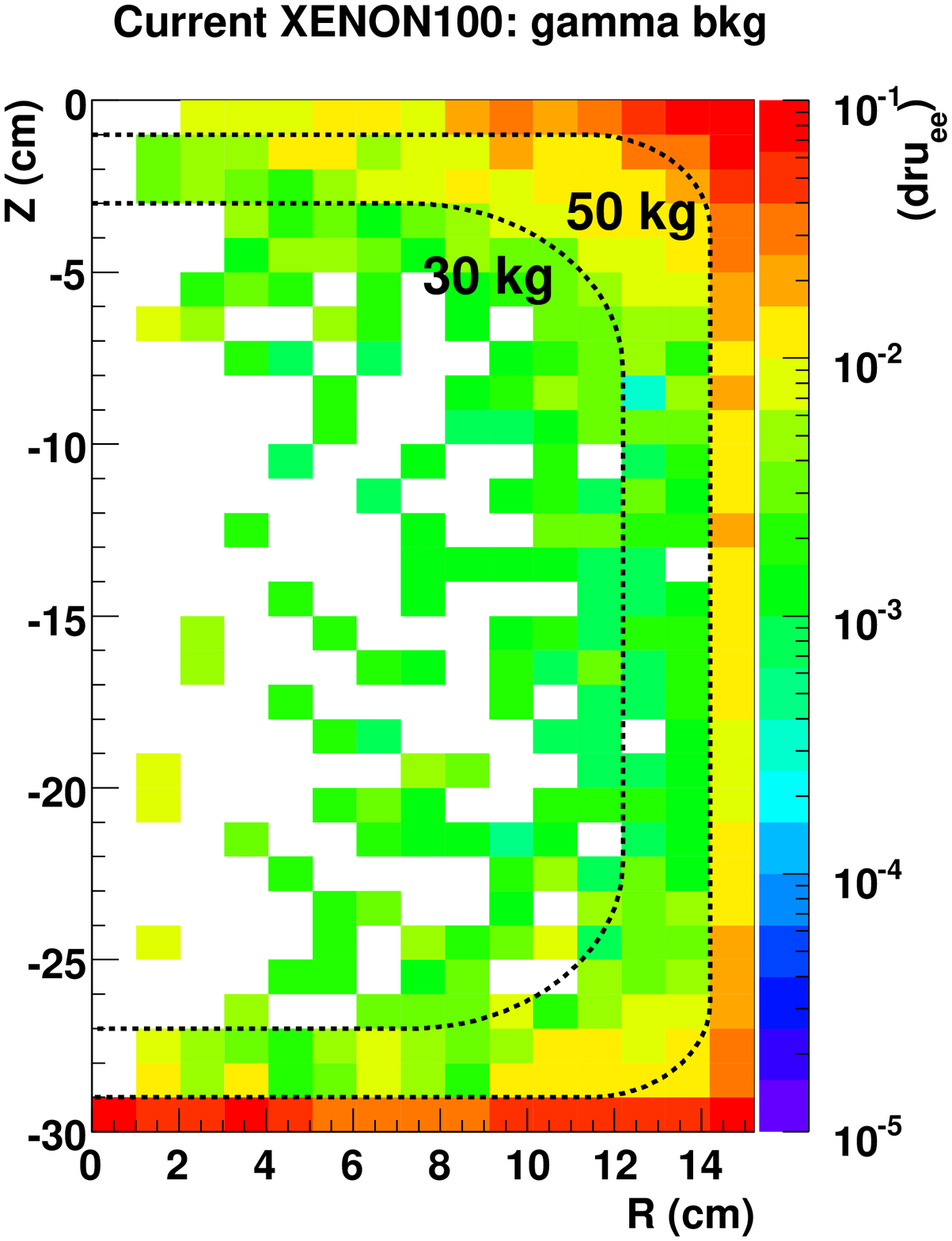} 
\includegraphics[scale=0.3]{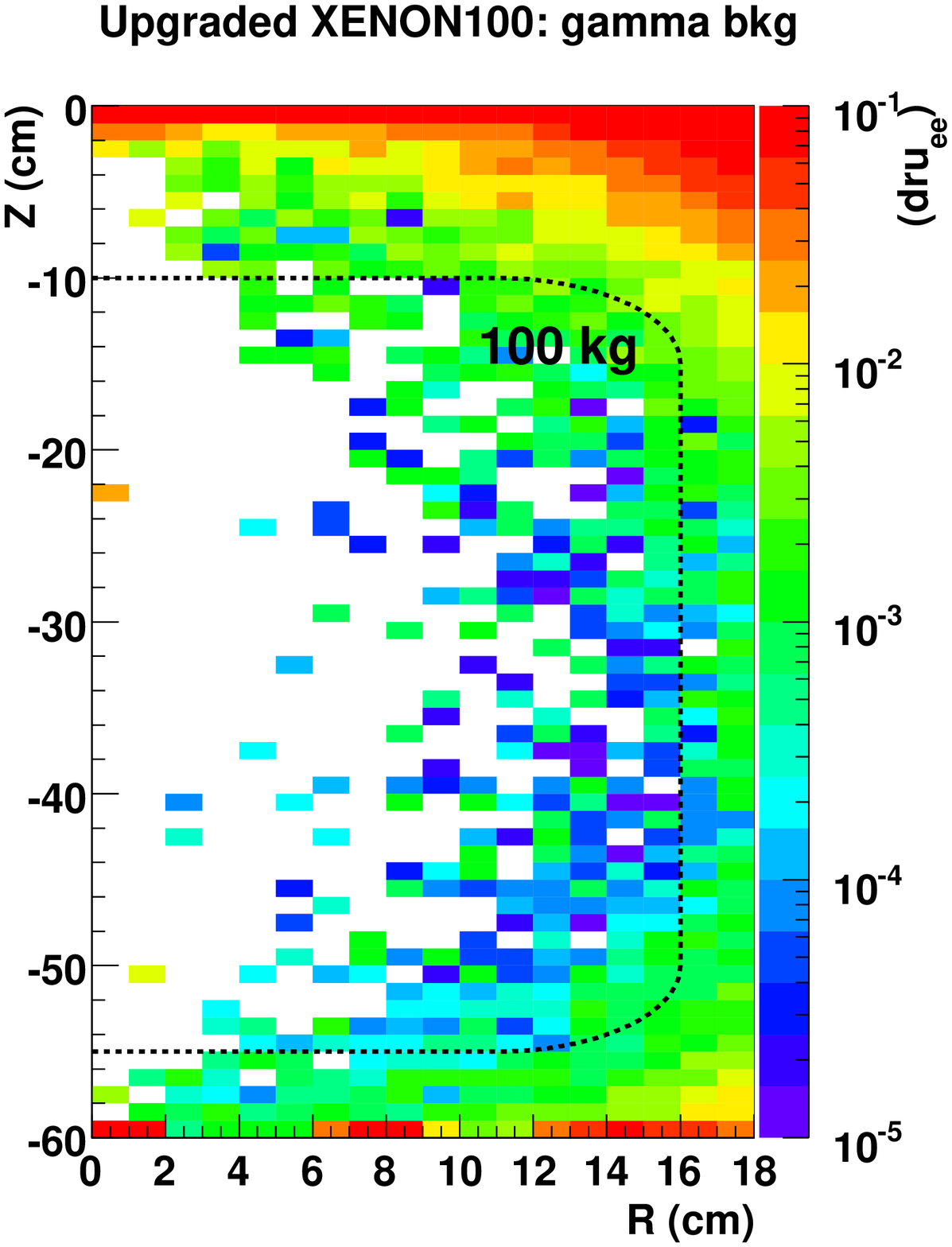} 
\caption[]{\small \label{bkg:2d}ER background rate in the current (left) and the upgraded (right) XENON100 detector. The dashed lines indicate the fiducial regions planned for the WIMP searches.}
\end{figure}

The electron recoil background is dominated by gamma rays from the decay chains of radioactive contaminants, mostly $^{238}$U, $^{232}$Th, $^{40}$K, 
and $^{60}$Co in the detector materials.  Using the measured activities of all materials, as listed in Table~\ref{tab:screen_results}, we have estimated 
the background rate with Monte Carlo simulations using Geant4. For the current XENON100 the background is dominated by the PMTs, as shown in Table~\ref{tab:bkg}.  For the upgraded detector,  the background rate  will be significantly reduced by  using QUPIDs and  a copper cryostat. Additionally, the self-shielding of LXe decreases the ER rate significantly in the central part of the target, as shown in Figure~\ref{bkg:2d}. The intrinsic ER rate due to $\beta$ decays of $^{85}$Kr will be reduced to $10^{-4}$ events/keV/kg/day, with a Kr/Xe level of 5\,ppt, achieved by using the cryogenic distillation column procured for the current experiment. 

$(\alpha,n)$ and spontaneous fission reactions in the detector, shielding materials, and surrounding  rock/concrete are the dominant sources of neutrons.  Based on the measured U/Th activities of  XENON100 materials (Table~\ref{tab:screen_results}), we calculated the neutron yield using the modified SOURCES4A code \cite{sources4A} and
simulated the nuclear recoil rates in the LXe target (Table~\ref{tab:bkg}). The dominant neutron flux comes from the rock/concrete walls of the underground laboratory. Since the predicted NR rate is higher than the one from the PMTs, it will be 
reduced by a factor of 10 by adding 20\,cm of polyethylene outside the current Pb shield. 

The muon flux at the 3100 mwe Gran Sasso depth is 22~m$^{-2}$day$^{-1}$~\cite{Mei:2005gm}. Neutrons generated by muons in the shield, due to electromagnetic and hadronic showers and direct spallation, yield a total NR background at the level of 2.7$\pm$0.7 single NRs/year in a 100\,kg target. To reduce this background, a 98\% efficient muon veto will be used  for the upgraded XENON100 experiment. The high energy muons interacting in the rock can produce highly penetrating neutrons with energies extending to few GeV, with an estimated single scatter NR rate smaller than 0.3/year/100\,kg. 

\begin{table}
\centering
\begin{minipage}{1.0\textwidth}
\centering
{\footnotesize
\begin{tabular}{lcc|cc|cc}
\hline
\hline
   &  \multicolumn{4}{c}{Current XENON100} & \multicolumn{2}{c}{Upgraded XENON100} \\
Fiducial Mass & \multicolumn{2}{c}{50~kg}  &   \multicolumn{2}{c}{30~kg}   &  \multicolumn{2}{c}{100~kg} \\
\hline
Background &   ER  &  NR  &  ER &  NR &  ER  &  NR   \\
Units \footnote{$dru_{ee}$ = evts/kg/keVee/day, $dru_{nr}$ = evts/kg/keVnr/day}             &  [$10^{-3} dru_{ee}$] & [$10^{-7} dru_{nr}$] & [$10^{-3} dru_{ee}$] &  [$10^{-7} dru_{nr}$]  & [$10^{-3} dru_{ee}$] & [$10^{-7} dru_{nr}$] \\
              \hline
\hline
PMTs and bases  & 4.91 &  3.25 & $<$1.4   & 2.87 & 0.098 & 0.23\\
QUPIDs  &  --  &  --  &  -- &  -- & $<$0.027 & $<$0.10 \\
Stainless steel & $<$2.01  & $<$2.01  & $<$0.35 & $<$1.66 & $<$0.052 & $<$0.14\\
PTFE & $<$0.18  & $<$6.99 & $<$0.03 & $<$5.04 & $<$0.017 & $<$1.60 \\
Copper Cryostat & -- & --  & -- & -- & $<$0.033 & $<$0.02\\
Polyethylene 	& $<$2.50 & $<$5.37 & $<$1.2 & $<$4.73 & $<$0.105  & $<$0.60\\
$^{85}$Kr/U/Th \footnote{with $<$ 5~ppt Kr/Xe and U/Th in Xe below $10^{-13}$ g/g. XENON100 upgrade requires another factor 10 reduction} \ & $<$0.2  & -- & $<$0.2  & -- & $<$0.02 & --\\
Concrete/Rocks\footnote{with an additional layer of 20~cm polyethylene outside the current shield}  &  --  &   1.34  & -- & 1.11  & -- & 0.2\\
$\mu$-induced n in shield  &    --   &    33   &  --  & 33  &  -- & 0.7\footnote{with a 98\% efficient muon veto} \\
$\mu$-induced n in rock & -- & $<3.7$  & -- & $<3.7$ & -- & $<$3.7 \\
\hline
Total Bkg  & $<$9.8 & $<$55.7 & $<$3.2 & $<$52.1 & $<$0.35 & $<$7.3\\
\hline
Run Time &  \multicolumn{2}{c}{40 days}  & \multicolumn{2}{c}{200 days}  & \multicolumn{2}{c}{600 days}   \\
Raw Exposure & \multicolumn{2}{c}{2000 kg-day} & \multicolumn{2}{c}{6000 kg-day} & \multicolumn{2}{c}{60000 kg-day} \\
Total Bkg events & \multicolumn{2}{c}{$<$1.0} & \multicolumn{2}{c}{$<$1.2} & \multicolumn{2}{c}{$<$1.4}\\
\# of WIMP events\footnote{for 100 GeV/$c^2$ WIMPs with spin-independent $\sigma_{\chi-p}=10^{-44}$~cm$^2$} & \multicolumn{2}{c}{3.9}  &   \multicolumn{2}{c}{11.8}      &  \multicolumn{2}{c}{118}  \\
SI $\sigma_{\chi-p}$ reach & \multicolumn{2}{c}{$6\times10^{-45}$~cm$^2$ }   & \multicolumn{2}{c}{$2\times10^{-45}$~cm$^2$ }   & \multicolumn{2}{c}{$2\times10^{-46}$~cm$^2$ } \\
\hline
\hline
\end{tabular}
}
\hfill
\end{minipage}
\caption{\small Predicted single scatter electron recoil (ER) and nuclear recoil (NR) background rates in the WIMP search region (2$-$12 keVee or 4.5$-$26.9 keVnr) in the central 50~kg (30~kg)  target for the current XENON100 and in the central 100~kg target for the upgraded XENON100. To estimate the total background, the number of WIMP events and the sensitivity reach, we assume 99.5\% ER rejection, 50\% NR acceptance and 90\% software efficiency.}
\label{tab:bkg}
\end{table}

\subsection{WIMP Search Plan and Expected Sensitivity}

\begin{figure}
\centering
\includegraphics[scale=0.4]{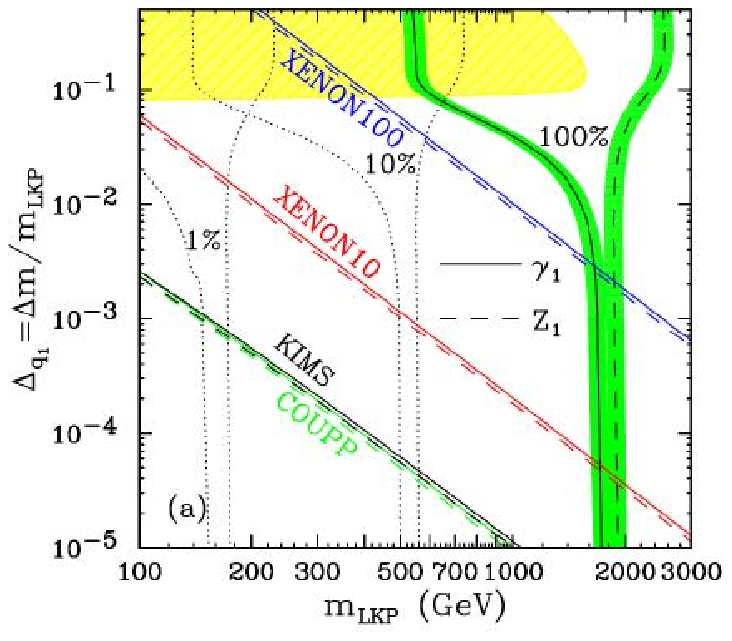} 
\includegraphics[scale=0.39]{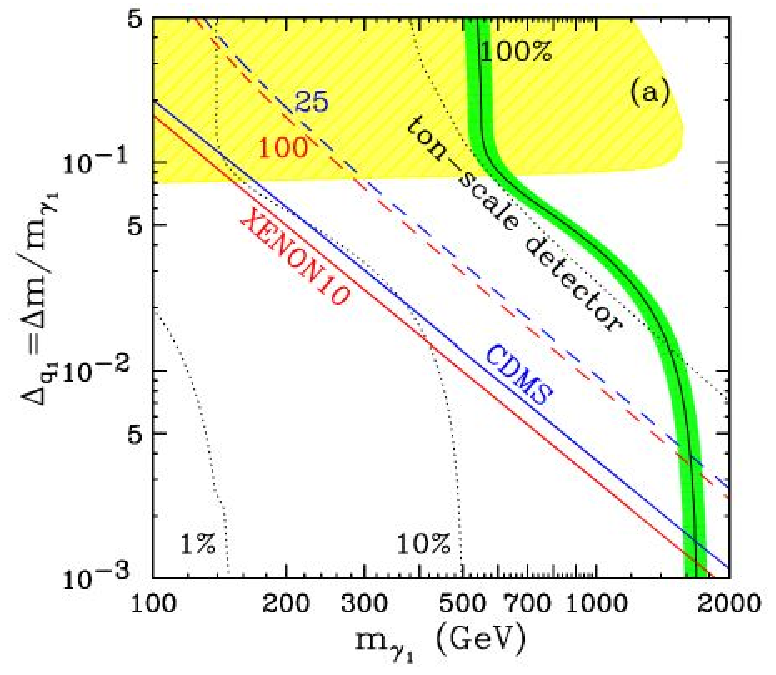} 
\caption[]{\small \label{fig:ued}Combined plot of  direct detection limits,  relic abundance constraints and the LHC reach for  $\gamma_1$ (SI and SD couplings) and  $Z_1$ (SD couplings),  in the parameter plane of the mass splitting $\Delta_{q_1}$  versus LKP mass ~\cite{arrenberg:prd08}. The SM Higgs mass is $m_h=120$ GeV. The black solid line accounts for all the dark matter (100\%) and the two black dotted lines show 10\% and 1\%, respectively. The green band shows the WMAP range, $0.1037 < \Omega_{CDM}h^2 < 0.1161$. The red solid lines show the current XENON10 limit, while the lines labeled 100, and XENON100, respectively, show the XENON100 projected sensitivities. The yellow region in the case of $\gamma_1$ LKP shows the parameter space which could be covered at the LHC in the $4\ell+\met$ channel, with a luminosity of 100 fb$^{-1}$ \cite{CFM}.}
\end{figure}

The background contribution from all expected sources is summarized in Table~\ref{tab:bkg} for the current and the upgraded XENON100 experiments. A WIMP search in the current XENON100 will be first performed in  a 50\,kg fiducial  target 
with 40 live-days of exposure, to reach a spin-independent WIMP sensitivity of $\sigma\sim 6\times10^{-45}$~cm$^2$ for 100 GeV/$c^2$ WIMPs. A factor of three improvement can be achieved after an additional 6 months of data taking and searching for WIMPs in a  30\,kg fiducial target,  to remain background free. The upgraded XENON100, with a lower background and larger target mass,  will be able to accumulate WIMP search data for two years in a 100\,kg fiducial target, reaching a sensitivity of $\sigma\sim 2\times10^{-46}$~cm$^2$ (see also Figure~\ref{fig:exclusion}). The XENON100 experiments will  thus probe a large fraction of  theoretically interesting SUSY models for both spin-dependent and spin-independent interactions.  They will also test other beyond standard model theories providing a dark matter candidate, such as universal extra dimensions. This is shown in Figure~\ref{fig:ued} for spin-dependent (left) and spin-independent (right) WIMP-nucleon 
couplings (Figures from ~\cite{arrenberg:prd08}). 

\end{document}